\documentclass[twocolumn,secnumarabic,amssymb, amsmath,nobibnotes, aps, prl,superscriptaddress,showpacs]{revtex4}
\begin{document}
\bibliographystyle{try}
\topmargin 0.05cm

%%%%%%%%%%%%%%%%%%%%%%%%%%%%%%%%%  TITLE %%%%%%%%%%%%%%%%%%%%%%%%%%%%%%%%
\title{Observation of an Exotic Baryon with $S=+1$
in Photoproduction from the Proton}
\newcommand*{\RPI }{ Rensselaer Polytechnic Institute, Troy, New York 12180-3590} 
\affiliation{\RPI } 

\newcommand*{\VANDER }{ Vanderbilt University, Nashville, Tennessee 37235}
\affiliation{\VANDER } 

\newcommand*{\JLAB }{ Thomas Jefferson National Accelerator Facility, Newport News, Virginia 23606} 
\affiliation{\JLAB } 

\newcommand*{\ASU }{ Arizona State University, Tempe, Arizona 85287-1504} 
\affiliation{\ASU } 

\newcommand*{\SACLAY }{ CEA-Saclay, Service de Physique Nucl\'eaire, F91191 Gif-sur-Yvette, Cedex, France} 
\affiliation{\SACLAY } 

\newcommand*{\UCLA }{ University of California at Los Angeles, Los Angeles, California  90095-1547} 
\affiliation{\UCLA } 

\newcommand*{\CMU }{ Carnegie Mellon University, Pittsburgh, Pennsylvania 15213} 
\affiliation{\CMU } 

\newcommand*{\CUA }{ Catholic University of America, Washington, D.C. 20064} 
\affiliation{\CUA } 

\newcommand*{\CNU }{ Christopher Newport University, Newport News, Virginia 23606} 
\affiliation{\CNU } 

\newcommand*{\UCONN }{ University of Connecticut, Storrs, Connecticut 06269} 
\affiliation{\UCONN } 

\newcommand*{\DUKE }{ Duke University, Durham, North Carolina 27708-0305} 
\affiliation{\DUKE } 

\newcommand*{\ECOSSEE }{ Edinburgh University, Edinburgh EH9 3JZ, United Kingdom} 
\affiliation{\ECOSSEE } 

\newcommand*{\FIU }{ Florida International University, Miami, Florida 33199} 
\affiliation{\FIU } 

\newcommand*{\FSU }{ Florida State University, Tallahassee, Florida 32306} 
\affiliation{\FSU } 

\newcommand*{\GWU }{ The George Washington University, Washington, DC 20052} 
\affiliation{\GWU } 

\newcommand*{\ECOSSEG }{ University of Glasgow, Glasgow G12 8QQ, United Kingdom} 
\affiliation{\ECOSSEG } 

\newcommand*{\INFNFR }{ INFN, Laboratori Nazionali di Frascati, Frascati, Italy} 
\affiliation{\INFNFR } 

\newcommand*{\INFNGE }{ INFN, Sezione di Genova, 16146 Genova, Italy} 
\affiliation{\INFNGE } 

\newcommand*{\ORSAY }{ Institut de Physique Nucleaire ORSAY, Orsay, France} 
\affiliation{\ORSAY } 

%\newcommand*{\BONN }{ Institute f\"{u}r Strahlen und Kernphysik, Universit\"{a}t Bonn, Germany} 
%\affiliation{\BONN } 

\newcommand*{\ITEP }{ Institute of Theoretical and Experimental Physics, Moscow, 117259, Russia} 
\affiliation{\ITEP } 

\newcommand*{\JMU }{ James Madison University, Harrisonburg, Virginia 22807} 
\affiliation{\JMU } 

\newcommand*{\KYUNGPOOK }{ Kyungpook National University, Daegu 702-701, South Korea} 
\affiliation{\KYUNGPOOK } 

\newcommand*{\MIT }{ Massachusetts Institute of Technology, Cambridge, Massachusetts  02139-4307} 
\affiliation{\MIT } 

\newcommand*{\UMASS }{ University of Massachusetts, Amherst, Massachusetts  01003} 
\affiliation{\UMASS } 

\newcommand*{\UNH }{ University of New Hampshire, Durham, New Hampshire 03824-3568} 
\affiliation{\UNH } 

\newcommand*{\NSU }{ Norfolk State University, Norfolk, Virginia 23504} 
\affiliation{\NSU } 

\newcommand*{\OHIOU }{ Ohio University, Athens, Ohio  45701} 
\affiliation{\OHIOU } 

\newcommand*{\ODU }{ Old Dominion University, Norfolk, Virginia 23529} 
\affiliation{\ODU } 

\newcommand*{\PITT }{ University of Pittsburgh, Pittsburgh, Pennsylvania 15260} 
\affiliation{\PITT } 

%\newcommand*{\ROMA }{ Universita' di ROMA III, 00146 Roma, Italy} 
%\affiliation{\ROMA } 

\newcommand*{\RICE }{ Rice University, Houston, Texas 77005-1892} 
\affiliation{\RICE } 

\newcommand*{\URICH }{ University of Richmond, Richmond, Virginia 23173} 
\affiliation{\URICH } 

\newcommand*{\SCAROLINA }{ University of South Carolina, Columbia, South Carolina 29208} 
\affiliation{\SCAROLINA } 

\newcommand*{\UTEP }{ University of Texas at El Paso, El Paso, Texas 79968} 
\affiliation{\UTEP }

\newcommand*{\UNIONC }{ Union College, Schenectady, NY 12308} 
\affiliation{\UNIONC } 

\newcommand*{\VT }{ Virginia Polytechnic Institute and State University, Blacksburg, Virginia   24061-0435} 
\affiliation{\VT } 

\newcommand*{\VIRGINIA }{ University of Virginia, Charlottesville, Virginia 22901} 
\affiliation{\VIRGINIA } 

\newcommand*{\WM }{ College of William and Mary, Williamsburg, Virginia 23187-8795} 
\affiliation{\WM } 

\newcommand*{\YEREVAN }{ Yerevan Physics Institute, 375036 Yerevan, Armenia} 
\affiliation{\YEREVAN }

%\newcommand*{\NONE }{ unknown} 
%\affiliation{\NONE } 

\newcommand*{\NOWNCATU }{ North Carolina Agricultural and Technical State University, Greensboro, NC 27411}

\newcommand*{\NOWECOSSEG }{ University of Glasgow, Glasgow G12 8QQ, United Kingdom}

\newcommand*{\NOWJLAB }{ Thomas Jefferson National Accelerator Facility, Newport News, Virginia 23606}

\newcommand*{\NOWSCAROLINA }{ University of South Carolina, Columbia, South Carolina 29208}

\newcommand*{\NOWFIU }{ Florida International University, Miami, Florida 33199}

\newcommand*{\NOWOHIOU }{ Ohio University, Athens, Ohio  45701}

\newcommand*{\NOWCMU }{ Carnegie Mellon University, Pittsburgh, Pennsylvania 15213}

\newcommand*{\NOWINDSTRA }{ Systems Planning and Analysis, Alexandria, Virginia 22311}

\newcommand*{\NOWASU }{ Arizona State University, Tempe, Arizona 85287-1504}

\newcommand*{\NOWCISCO }{ Cisco, Washington, DC 20052}

\newcommand*{\NOWUK }{ University of Kentucky, LEXINGTON, KENTUCKY 40506}

\newcommand*{\NOWINFNFR }{ INFN, Laboratori Nazionali di Frascati, Frascati, Italy}

\newcommand*{\NOWUNCW }{ North Carolina}

\newcommand*{\NOWHAMPTON }{ Hampton University, Hampton, VA 23668}

\newcommand*{\NOWTulane }{ Tulane University, New Orleans, Lousiana  70118}

\newcommand*{\NOWKYUNGPOOK }{ Kungpook National University, Taegu 702-701, South Korea}

\newcommand*{\NOWCUA }{ Catholic University of America, Washington, D.C. 20064}

\newcommand*{\NOWGEORGETOWN }{ Georgetown University, Washington, DC 20057}

\newcommand*{\NOWJMU }{ James Madison University, Harrisonburg, Virginia 22807}

\newcommand*{\NOWCALTECH }{ California Institute of Technology, Pasadena, California 91125}

\newcommand*{\NOWMOSCOW }{ Moscow State University, General Nuclear Physics Institute, 119899 Moscow, Russia}

\newcommand*{\NOWVIRGINIA }{ University of Virginia, Charlottesville, Virginia 22901}

\newcommand*{\NOWYEREVAN }{ Yerevan Physics Institute, 375036 Yerevan, Armenia}

\newcommand*{\NOWRICE }{ Rice University, Houston, Texas 77005-1892}

\newcommand*{\NOWBATES }{ MIT-Bates Linear Accelerator Center, Middleton, MA 01949}

\newcommand*{\NOWODU }{ Old Dominion University, Norfolk, Virginia 23529}

\newcommand*{\NOWVSU }{ Virginia State University, Petersburg,Virginia 23806}

\newcommand*{\NOWORST }{ Oregon State University, Corvallis, Oregon 97331-6507}

\newcommand*{\NOWGWU }{ The George Washington University, Washington, DC 20052}

\newcommand*{\NOWMIT }{ Massachusetts Institute of Technology, Cambridge, Massachusetts  02139-4307}

\newcommand*{\NOWINFNGE }{ INFN, Sezione di Genova, 16146 Genova, Italy}

\newcommand*{\NOWdeceased }{ Deceased}

%%%%%%%%%%%%%%%%%%%% authors %%%%%%%%% 
%vpk,lg,dw,paul stoler, m. battaglieri, r. devita,g. adams, j.li,m.nozar,c.salgado  
\author{V.~Kubarovsky}
     \affiliation{\RPI}\affiliation{\JLAB}
\author{L.~Guo}
     \affiliation{\VANDER}
\author{D.P.~Weygand}
     \affiliation{\JLAB}
\author{P.~Stoler}
     \affiliation{\RPI}
\author{M.~Battaglieri}
     \affiliation{\INFNGE}
\author{R.~DeVita}
     \affiliation{\INFNGE}
\author{G.~Adams}
     \affiliation{\RPI}
\author{Ji~Li}
     \affiliation{\RPI}
\author{M.~Nozar}
     \affiliation{\JLAB}
     \altaffiliation{\NONE}
\author{C.~Salgado}
     \affiliation{\NSU}
% ...

\collaboration{The CLAS Collaboration}
     \noaffiliation

\date{\today}

\begin{abstract}                % DON'T CHANGE THIS LINE

The reaction $\gamma p \rightarrow \pi^+K^-K^+n$ was studied 
at Jefferson Lab using a tagged photon beam with an energy range of 3-5.47~GeV.
A narrow baryon state with strangeness S=+1 and 
mass $M=1555\pm 10$~MeV/c$^2$ was observed in the  $nK^+$ invariant  
mass spectrum. The peak's width is consistent with the CLAS resolution
(FWHM=26~MeV/c$^2$),
and its statistical significance is 7.8 $\pm$ 1.0 ~$\sigma$.
A baryon  with positive strangeness has  exotic 
structure and
cannot be described in the framework of the naive 
constituent quark model.
The mass of the observed state is consistent with the mass predicted 
by the chiral soliton model for the $\Theta^+$ baryon.
In addition, the $pK^+$ invariant mass distribution was analyzed in the
reaction $\gamma p\rightarrow K^-K^+p$ with high statistics in search of doubly-charged exotic baryon states.
No resonance structures were found in this spectrum.

\end{abstract}
\maketitle
\newpage

The constituent quark model 
describes light mesons as  bound states of a quark and
an antiquark ($q\bar q$), and baryons  as  bound states  of
three quarks ($qqq$),
 where $q$ is $u$, $d$ or $s$.
It is a remarkable feature of meson 
and baryon spectroscopy
that practically all well-established particles can be categorized using
this naive  model. At the same time, Quantum Chromodynamics (QCD)
predicts the existence of so-called exotic mesons and baryons with  more
complicated internal  structures. Exotic mesons may be classified such as
glueballs ($ggg$), hybrids
($q\bar qg$), and four--quark
($q\bar qq\bar q$) states, and  
exotic baryons as  $(qqqg)$ or $(qqqq\bar q)$ states. 

A baryon with strangeness quantum number $S=+1$
is an excellent example of a particle whose exotic structure is manifest.
Diakonov, Petrov, and Polyakov \cite{Diakonov}, in the framework of the
chiral soliton model, have predicted a spin $1/2$, isospin $0$, and
strangeness $S=+1$ exotic baryon $\Theta^+$ with mass  $M \sim 1.53$
GeV/c$^2$.
Also, the pentaquark states was considered in the quark models
\cite{jaffe} and lattice QCD \cite{fodor}.
The possible quark structure of $\Theta^+$ is $(uudd\bar s)$.

Experimental evidence for a narrow $S=+1$ baryon state has been reported 
in the interactions of photons and kaons with nuclear targets:
$\gamma n \rightarrow K^+K^-n$ (on a $^{12}C$ target) \cite{Nakano},
$K^+Xe$ collisions 
in the $pK^0$  decay mode \cite{Dolgolenko}, and the exclusive photo-deuteron 
interaction
in the $nK^+$  decay mode \cite{Ken}. 
In all these reactions the primary interaction is with 
a neutron in the initial state.
The CLAS collaboration has reported the observation of the same state
in the photoproduction reaction from the proton target \cite{vpk}.
The SAPHIR collaboration
 \cite{SAPHIR} has also reported the observation of this state in
the reaction $\gamma p \rightarrow nK^+K^0_s$.
A narrow peak with mass 
around $M=1.54$ GeV/c$^2$ and width less than $25$~ MeV/c$^2$ 
was observed, and interpreted as the production 
and decay of the  exotic $\Theta^+$ baryon. More recently, the ITEP
collaboration 
observed a narrow baryon resonance in the invariant mass of the $p
K_{S}^0$ system 
formed in neutrino and antineutrino collisions with nuclei \cite{itep}.

This letter reports a more comprehensive study than in \cite{vpk} of the $\Theta^+$
production on a proton target which includes data from three distinct runs under
different experimental 
conditions in CLAS \cite{CLAS}. Two reactions, 
$\gamma p \rightarrow \pi^+ K^+K^- n$ and
$\gamma p \rightarrow K^+K^- p$, have been analyzed. Of the three
runs, $Run\ a$ 
and $Run\ b$ had identical geometrical acceptance and 
trigger requirement with $Run\ c$ having slightly different running conditions.
$Runs\ a, b$ and $c$ had a tagged 
photon beam in the energy range of 3.2--3.95~GeV, 
3--5.25~GeV, and 
4.8--5.47~ GeV, respectively \cite{TAGGER}. CLAS is a six-fold
segmented toroidal 
magnetic spectrometer (details described in \cite{CLAS}).
$Run\ c$ triggered on the events with at least 2 out of 6 CLAS
sectors having charged tracks, while $Runs\ a$ and $b$ triggered on events with hits in opposite
sectors. $Runs\ a$ and $b$ had the hydrogen target in the 
standard position, but in  $Run\ c$ the target was moved upstream   by 1 
meter to improve the CLAS acceptance in the forward direction,
especially for the 
negative charged particles.   
The estimated integrated luminosity for the combined data of $Runs\ a$
and $b$ is 
2 pb$^{-1}$, and $Run\ c$ is 2.7 pb$^{-1}$. The combined analysis of
these three 
runs offers access to a wider range of acceptance and energies. 

\begin{figure}[ht]
\vspace{80mm}
{\includegraphics{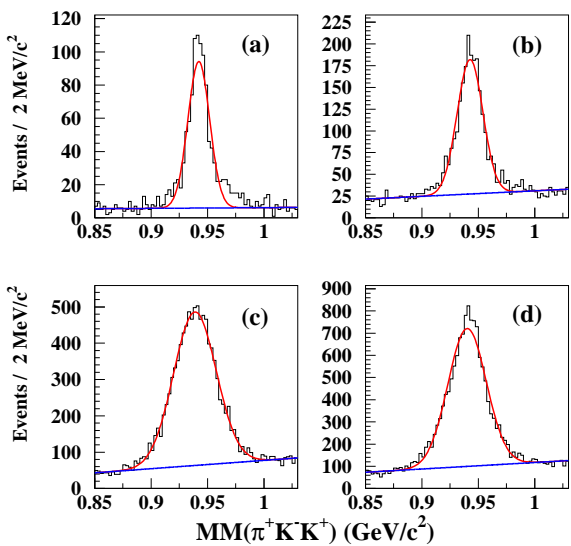}}
\caption{ 
The missing mass distributions in the reaction $\gamma p\rightarrow \pi^+K^-K^+X$
for the three different runs ($Runs$\ $a$,$b$, and $c$) and the combined data spectrum ($d$).
The mass resolutions are 10, 11, and 19 MeV/c$^2$ for $Runs$\ $a$, $b$, and $c$  respectively.
}
\label{neutron}
\end{figure}

Events having a
$\pi^+,K^+$, and $K^-$
in the final state, identified by time-of-flight,   were selected for the analysis
of the reaction
$\gamma p\rightarrow \pi^+K^-K^+n$.
The missing mass distributions for the reaction 
$\gamma p\rightarrow \pi^+K^-K^+X$
are shown in  Fig. \ref{neutron}.
%%%%%%%%%%%%%%%%%%%%%%%%%% FIG: neutron %%%%%%%%%%%%%%%%%%%%%%%%%%%
%%%%%%%%%%%%%%%%%%%%%%%%%%%%%%%%%%%%%%%%%%%%%%%%%%%%%%%%%%%%%%%%%%%%%%
A neutron  peak is clearly seen in each of
these distributions. The fitted masses of the neutron peaks
are consistent with each other within 3 MeV/c$^2$.
The mass resolution for $Runs$\ $a$ and $b$ is about twice as good 
as in $Run\ c$ due to the fact that $Run\ c$ had higher energy and
the magnetic field was
reduced by a factor of 2.
Events within  $\pm 2\sigma$ of the neutron peak
were retained for each run individually, 
resulting in a total of 14k events.

\begin{figure}[ht]
\vspace{80mm}
{\includegraphics{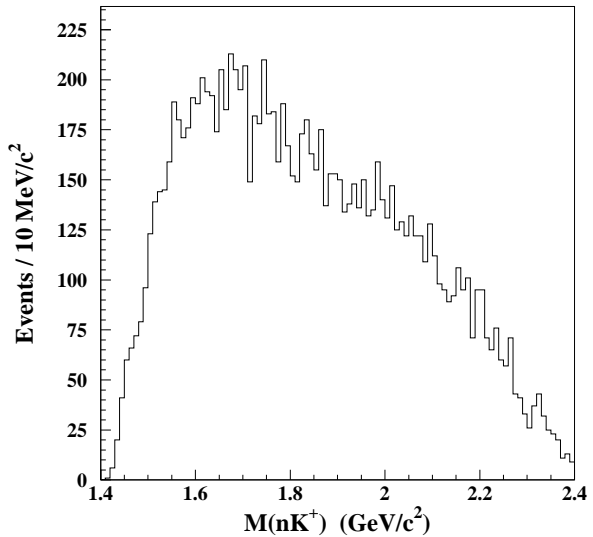}}
\caption{ 
The $n K^+$ invariant mass spectrum  
in the reaction 
$\gamma p\rightarrow \pi^+K^-K^+(n)$. 
The neutron was measured from the missing four-momentum. 
}
\label{theta_1}
\end{figure}

There are about 200  $\phi$ mesons  in the selected sample, which were
removed 
by eliminating events with   the 
$K^+ K^-$ effective mass less than 1.06~GeV/c$^2$. The final $n K^+$
invariant 
mass spectrum calculated from missing mass 
in the reaction  
$\gamma p\rightarrow \pi^+ K^- X$, combining data from all three data runs, 
is shown  in  Fig. \ref{theta_1}. No obvious structure is seen in this spectrum.

%%%%%%%%%%%%%%%%%%%%%%%%%%% FIG: nK+  %%%%%%%%%%%%%%%
%%%%%%%%%%%%%%%%%%%%%%%%%%%%%%%%%%%%%%%%%%%%%%%%%%%%%%%%%%%

%%%%%%%%%%%%%%%%%%%%%%%%%%% FIG: feyman diagrams %%%%%%%%%%%%%%%
\begin{figure}[ht]
\vspace{80mm}
{\includegraphics{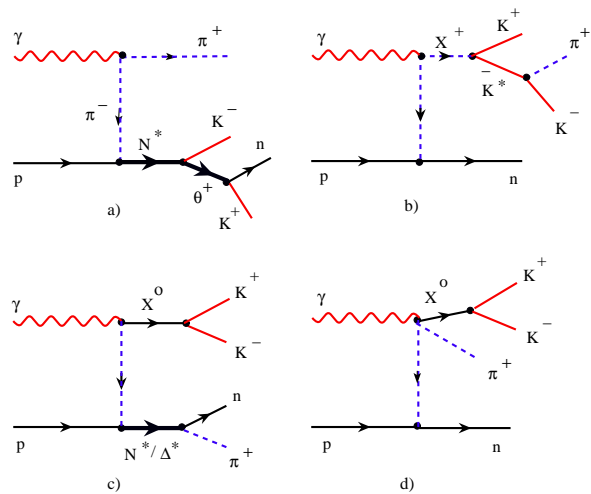}}
\caption{ 
The diagram in ($a$) shows a possible production mechanism for the
$\Theta^+$, 
which could be a decay product of an intermediate baryon
resonance. The three 
diagrams in ($b$), ($c$), and ($d$) are  background processes in the
reaction 
$\gamma p\rightarrow \pi^+K^-K^+(n)$. All background processes have a
$K^+$ 
going in the forward direction in the center-of-mass system.}
\label{dia}
\end{figure}
%%%%%%%%%%%%%%%%%%%%%%%%%%%%%%%%%%%%%%%%%%%%%%%%%%%%%%%%%%%%%%%

%%%%%%%%%%%%%%%%%%%%%%%%%%% FIG: nK+ angle cut %%%%%%%%%%%%%%%
\begin{figure}[ht]
\vspace{80mm}
{\includegraphics{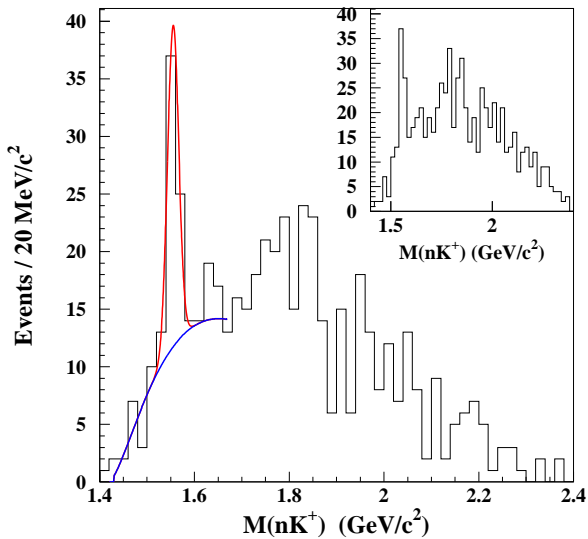}}
\caption{ 
The $nK^+$ invariant mass spectrum  
in the
reaction $\gamma p\rightarrow \pi^+K^-K^+(n)$ with the cut
$\cos\theta^*_{\pi^+}>0.8$ and $\cos\theta^*_{K^+}<0.6$.
$\theta^*_{\pi^+}$ and $\theta^*_{K^+}$ are the
angles between the $\pi^+$ and $K^+$ mesons
and photon beam in the center-of-mass system. 
The background function we used in the fit was obtained from the simulation.
The inset shows the $nK^+$ invariant mass spectrum  with only the $\cos\theta^*_{\pi^+}>0.8$ cut.
}
\label{theta_2}
\end{figure}
%%%%%%%%%%%%%%%%%%%%%%%%%%%%%%%%%%%%%%%%%%%%%%%%%%%%%%%%%%%%%%%

We explored various possible t-channel processes  to understand the  potential production mechanisms for the $\Theta^+$
as well as the background, examples of which are illustrated in  Fig. \ref{dia}.
A peak  appears most clearly when
requiring  $\cos\theta^*_{\pi^+}>0.8$, where
$\theta^*_{\pi^+}$ is the center-of-mass angle between the $\pi^+$ and the
photon beam. This requirement approximately 
corresponds to  $-t<0.28$ GeV/c$^2$ where 
$t=(k-p)^2$, $k$ is the photon 4-momentum, and $p$ is the pion 4-momentum.
This would correspond to an enhancement of the t-channel process
as shown in Fig.~\ref{dia}~$a$. The spectrum is shown in the inset 
in Fig.~\ref{theta_2}. 
As a systematic check, 
we  varied the pion angular cut from 0.7 to 0.9 and found that 
in all cases the peak was clearly visible.

The background reaction  $\gamma p\rightarrow \pi^+K^-K^+n$ is dominated by 
meson resonance production decaying to $K^+ K^-$ with the excitation 
of baryon resonances decaying to 
$n \pi^+$, or meson resonance production decaying to $K^+ K^- \pi^+$, both
with small momentum transfer to the meson system. These processes have the $K^+$
moving forward in the center-of-mass system (Fig. \ref{dia}~$b,c$ and $d$). 
To suppress such backgrounds,
a cut was applied to eliminate  events having a positive kaon
going forward  with $\cos\theta^*_K>0.6$, where
$\theta^*_{K^+}$ is the center-of-mass angle between the $K^+$ and the
photon beam. 
 The remaining data sample is virtually free of the contaminating
events 
that have  baryons decaying to $n \pi^+$ in this final state since 
the $\pi^+$ from such event will most likely not move very forward in the center-of-mass system. 
The $\Theta^+$ peak was
clearly observed in each of the three data sets; 
the resulting $nK^+$ mass spectrum are combined 
and  shown in Fig. \ref{theta_2}. 

An investigation has been conducted to test whether a narrow peak in 
the $n K^+$ invariant mass spectrum can be artificially manufactured.
First, we checked the sidebands around the neutron in Fig.~\ref{neutron};
the resulting $n K^+$ effective mass distribution is structureless.
We also considered the effect of  
the kinematic requirements that we applied by performing   a Monte 
Carlo simulation based on  $n K^+ K^- \pi^+$ 4-body phase space, 
$n K^+ \bar{K^*_0}$ 3-body phase space, and 
t-channel meson production. The meson events in the latter process 
are generated using $K^+ K^- \pi^+$ 3-body phase space and the shape
 of the $K^+ K^- \pi^+$ invariant mass distribution from the data.  
We found no structure generated using the same cuts on the simulated 
events as applied on the data. 
Furthermore, one may consider whether some particular combination 
of meson waves can be reflected as a narrow peak in the $n K^+$ invariant mass distribution. 
We performed a full partial wave analysis of the $K^+ K^- \pi^+$ meson system
on $Run\  c$ data and utilized the prediction of this analysis to further 
probe the possibility of  meson reflection into the $n K^+$ invariant 
mass spectrum. Again, we found that with a set of meson partial waves
that 
well describes the entire data set we did not generate a narrow
$\Theta^+$ 
peak when the same angular cuts as above are applied. 
 
\begin{figure}[ht]
\vspace{80mm}
{\includegraphics{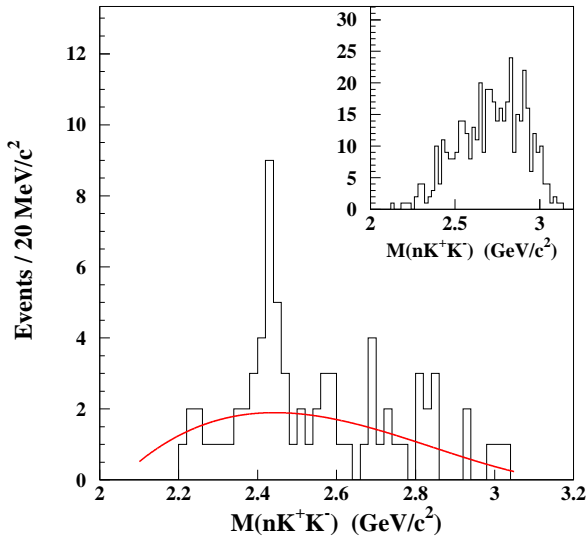}}
\caption{ 
The $n K^+ K^-$ invariant mass spectrum calculated from the missing mass off the $\pi^+$
in the
reaction $\gamma p\rightarrow \pi^+K^-K^+(n)$ with the cuts
$\cos\theta^*_{\pi^+}>0.8$ and $\cos\theta^*_{K^+}<0.6$.
$\theta^*_{\pi^+}$ and $\theta^*_{K^+}$ are the
 angles between the $\pi^+$ or $K^+$ mesons
and photon beam in the center-of-mass system. 
These events have $M(K^+n)$ between 
1.54 and 1.58 GeV/c$^2$. The shape of the background curve was
obtained 
from the simulation as discussed in the text. The inset shows the
$nK^+K^-$ 
invariant mass spectrum for all other events in Fig. \ref{theta_2}.
}
\label{nkk_1}
\end{figure}

The final $nK^+$ effective mass distribution (Fig. \ref{theta_2}) 
was fitted  by the sum of a Gaussian function and a 
 background function obtained from the simulation.
The fit parameters are: 
$N_{\Theta^+}=41\pm10$, 
$M=1555\pm 1$ MeV/c$^2$, and $\Gamma=26\pm7$ MeV/c$^2$ (FWHM), 
where the errors are statistical.
The systematic mass scale uncertainty is estimated to be $\pm 10$
MeV/c$^2$. 
This uncertainty is larger than our previously reported uncertainty
\cite{Ken} 
because of the different energy range and running conditions, and is
mainly 
due to the momentum calibration of the
CLAS detector and the photon beam energy calibration.
The statistical significance for the fit in  Fig.~\ref{theta_2}
over a $40$ MeV/c$^2$ mass window is calculated as
$N_{P}/\sqrt{N_{B}}$, 
where $N_{B}$ is the number of counts in the background fit under the
peak 
and  $N_{P}$ is the number of counts in the peak. We estimate the
significance 
to be 7.8 $\pm$ 1.0~$\sigma$. The uncertainty of 1.0 $\sigma$ is due
to 
the different background functions that we tried. When a simple
polynomial 
background is used, the statistical significance is higher. In the
present 
analysis we used the background function obtained from the simulation
as 
discussed above.
The fact that the angular cuts we applied enhanced the $\Theta^{+}$
signal  
suggests  the possible 
production of an $N^*/\Delta^*$ that decays to $\Theta^+$ 
and $K^-$. 
If the $\Theta^+$ is an isosinglet, the intermediate state can only be an $N^*$.
The $ n K^+ K^-$ invariant mass is shown in Fig. \ref{nkk_1} for the
events 
with $nK^+$ effective mass 
between 1.54 and 1.58 GeV/c$^2$.  The apparent
excess of events near $2.4$ GeV/c$^2$ is suggestive of an intermediate
 baryon state. A possible production mechanism that could contribute to 
the $\Theta^+$ production is shown in Fig.~\ref{dia}~$a$. Similar processes 
were also theoretically  considered in Ref.~\cite{philip}.
The simulation  we described previously also demonstrated that 
the angular cuts we applied could not generate a narrow peak in 
the $n K^{+} K^{-}$ invariant mass spectrum from any of the three data runs.

%%%%%%%%%%%%%%%%%%%%%%%%%%% FIG: nK+K- mass %%%%%%%%%%%%%%%
%%%%%%%%%%%%%%%%%%%%%%%%%%%%%%%%%%%%%%%%%%%%%%%%%%%%%%%%%%%%%%%

In addition,
a search for a manifestly exotic baryon 
($Q=2$, $S=+1$) was performed in the reaction
$\gamma p\rightarrow K^-X^{++},~X^{++}\rightarrow pK^+$.
There were 225k events
with a proton and $K^+$ in the
final state, which were selected for the analysis of this reaction.
The $K^-$ was identified by the missing mass technique.
After the removal of 
$\phi\to K^+K^-$ and $\Lambda(1520)\to K^-p$,
we observe no resonant structures  in
the $pK^+$ invariant mass distribution for the remaining 130k events.
The  $pK^+$ invariant mass spectra for
different $\cos\theta^*_{K^-}$ were analyzed as well, where
$\theta^*_{K^-}$ is the angle
between the $K^-$ and incident photon in the center-of-mass system.
There are no resonance structures evident in any of these distributions.
A more detailed analysis will be presented in a future paper.

In summary, 
the reaction $\gamma p \rightarrow \pi^+K^-K^+n$ was studied 
at Jefferson Lab with photon energies from 3  to 5.47~GeV 
using the CLAS detector.
A narrow baryon state with positive strangeness $S=+1$, mass 
$M=1555\pm 10$~MeV/c$^2$ and width 
$\Gamma<26$ MeV/c$^2$ (FWHM) was observed.
The width is close to the experimental mass resolution of the CLAS detector.
The peak's statistical significance is 7.8 $\pm$ 1.0~$\sigma$. 
In addition, the $pK^+$ invariant mass distribution was analyzed in the
reaction $\gamma p\rightarrow K^-K^+p$ with high statistics.
No resonance structures were found in this spectrum.

We would like to acknowledge the outstanding efforts of the staff of the 
Accelerator and the Physics Divisions at JLab that made this experiment possible.
This work was supported in part 
by 
the U.S. Department of Energy, 
the National Science Foundation, 
the Istituto Nazionale di Fisica Nucleare, 
the  French Centre National de la Recherche Scientifique, 
the French Commissariat \`{a} l'Energie Atomique, 
an Emmy Noether grant from the Deutsche Forschungs gemeinschaft, and 
the Korean Science and Engineering Foundation.
The Southeastern Universities Research Association (SURA) operates the 
Thomas Jefferson National Accelerator Facility for the United States 
Department of Energy under contract DE-AC05-84ER40150. 

%%%%%%%%%%%%%%%%%%%%%%%%%%%%%%%%%A

\section*{}

\end{document}